\documentclass[journal,comsoc]{IEEEtran}  

\IEEEoverridecommandlockouts                              

\usepackage{amsmath} 
\usepackage{amssymb}
\usepackage{bm}
\usepackage{array}
\usepackage{subfig}

\usepackage[english]{babel}
\usepackage[T1]{fontenc}
\usepackage{cite,url,color} 
\usepackage{graphics,amsfonts}
\usepackage{epstopdf}
\usepackage[pdftex]{graphicx}
\usepackage[table]{xcolor}
\usepackage[nomain,xindy,acronym,nonumberlist]{glossaries}

\definecolor{mycolor1}{rgb}{0.00000,0.44700,0.74100}%
\definecolor{mycolor2}{rgb}{0.85000,0.32500,0.09800}%
\definecolor{mycolor3}{rgb}{0.92900,0.69400,0.12500}%
\definecolor{mycolor4}{rgb}{0.49400,0.18400,0.55600}%
\definecolor{mycolor5}{rgb}{0.46600,0.67400,0.18800}%
\definecolor{mycolor6}{rgb}{0.30100,0.74500,0.93300}%

\definecolor{violet}{rgb}{0.6,0,0.6}%
\definecolor{orange_D}{rgb}{1.0000,0.5,0}%
\definecolor{cyan}{rgb}{0,0.67,0.64}%
\definecolor{red}{rgb}{0.9,0,0}%
\definecolor{green}{rgb}{0,0.8,0}%
\definecolor{gray}{rgb}{0.3,0.3,0.3}%

\def \fwidth{0.8\columnwidth}
\def \fheight {0.54\columnwidth}

\newacronym{kf}{KF}{Kalman Filter}
\newacronym{tkf}{TKF}{Tobit Kalman Filter}
\newacronym{ekf}{EKF}{Extended Kalman Filter}
\newacronym{ukf}{UKF}{Unscented Kalman Filter}
\newacronym{atkf}{ATKF}{Adaptive Tobit Kalman Filter}
\newacronym{akf}{AKF}{Adaptive Kalman Filter}
\newacronym{pdf}{PDF}{Probability Density Function}
\newacronym{cdf}{CDF}{Cumulative Distribution Function}
\newacronym{map}{MAP}{Maximum A Posteriori}
\newacronym{mle}{MLE}{Maximum Likelihood Estimator}
\newacronym{mse}{MSE}{Mean Squared Error}
\newacronym{als}{ALS}{Autocovariance Least Squares}
\newacronym{arima}{ARIMA}{Auto Regressive Integrated Moving Average}
\newacronym{ice}{ICE}{Innovation Covariance Estimator}
\newacronym{iot}{IoT}{Internet of Things}
\newacronym{uav}{UAV}{Unmanned Aerial Vehicle}
\newacronym{ble}{BLE}{Bluetooth Low Energy}
\newacronym{lora}{LoRaWAN}{Long Range Wide Area Network}
\newacronym{rss}{RSS}{Received Signal Strength}
\newacronym{aoa}{AoA}{Angle of Arrival}
\newacronym{vlc}{VLC}{Visible Light Communication}

\newcommand{\vast}{\bBigg@{4}}

\usepackage{placeins}

\usepackage{booktabs}
\usepackage{setspace}

\usepackage{pgfplots}
\usepackage{adjustbox}

\usepackage{gincltex}
\usepgfplotslibrary{fillbetween}

\title{
The Adaptive Tobit Kalman Filter: Tracking Position with Censored Measurements in the IoT
}

\author{\IEEEauthorblockN{Federico Chiariotti}

\IEEEauthorblockA{Department of Information Engineering, University of Padova\\Via Gradenigo, 6/b, 35131 Padova, Italy. Email: {\tt chiariot@dei.unipd.it}}
}
\begin{document}

\maketitle
\thispagestyle{empty}
\pagestyle{empty}

\begin{abstract}
In the \gls{iot} paradigm, distributed sensors and actuators can observe and act on their environment, communicating wirelessly. In this context, filtering the observations and tracking the network and environment state over time is extremely important, and the \gls{kf} is one of the most common tools for this.
Several of these applications deal with censored data, either because of sensor saturation or limited detection regions: when censoring happens, all measurements below a certain threshold are clipped to the threshold value. The recently proposed \gls{tkf} is an adjusted version of the \gls{kf} that can deal with censored measurements. However, like the traditional \gls{kf}, it needs full knowledge of the process and measurement noise covariances to work, which are not always available in practice. In this work, we relax this assumption and propose the \gls{atkf}, which can dynamically estimate the process and measurement noise along with the hidden state of the system from the censored measurements. We apply our solution to navigation and positioning \gls{iot} scenarios, obtaining a negligible performance loss with regard to the \gls{tkf}, even with no \emph{a priori} knowledge of the noise statistics.
\end{abstract}


\glsresetall

\section{Introduction}\label{sec:intro}
The \gls{iot} is a paradigm that exploits ubiquitous sensing by a myriad of low-power connected devices~\cite{gubbi2013internet}. Several \gls{iot} services aggregate sensor measurements and track important variables, such as the location of users and nodes~\cite{zhang2012energy}, over time. Filtering tools and estimators are commonly deployed in order to perform sensor data fusion~\cite{gite2016context} and correct for sensor error and drift~\cite{kumar2013automatic}. 

Localization and motion tracking are key aspects of the \gls{iot}~\cite{macagnano2014indoor}: several services need to be aware of users' and nodes' location to improve performance~\cite{jara2014mobile}, or even to work at all~\cite{perera2015energy}. The development of accurate filtering and tracking tools to predict the future position from past measurements~\cite{lin2016enhanced} can significantly improve these services' performance by introducing anticipatory elements~\cite{bui2017survey} in the optimization.

The \gls{kf}~\cite{kalman1960new} is the optimal estimator for linear dynamic systems, and it is now used in a wide range of \gls{iot} positioning and tracking applications, based on wireless signals~\cite{yu2017precise}, video object recognition~\cite{weng2006video} and vehicle GPS~\cite{dellaert1997robust}. The \gls{kf} exploits knowledge of the system model to estimate a hidden state from noisy measurements.
However, its performance degrades strongly when there are nonlinearities in the system or non-Gaussian noise distributions. One of the most common nonlinearities is Tobit Type I censoring~\cite{tobin1958estimation}: in this kind of model, measurements have a saturation threshold (that might be due to sensor saturation, detection limits, or occlusion), and any value below the threshold is clipped. It is easy to extend the model to the case with an upper threshold instead of a lower one, or even to the doubly censored case. The \gls{tkf}~\cite{allik2015tobit} is a recent adaptation of the standard \gls{kf} that can deal with this kind of nonlinearities.

However, an important assumption of the \gls{kf} is the full knowledge of the noise statistics: in order to correctly separate changes in the hidden state from measurement noise, the filter needs the covariance matrices of the process and measurement noise.

Leveraging the approaches proposed in~\cite{allik2015tobit} and~\cite{gao2015adaptive}, this work derives the \gls{atkf}: this filter can overcome both the censoring issue and the noise estimation problem, correctly estimating the state with no \emph{a priori} knowledge of the noise covariance even when most of the measurements are censored. First, the unbiased noise estimator from~\cite{gao2015adaptive} is adapted to obtain the one-step~\gls{mle} for the Tobit case. The \gls{atkf} is then tested in a simple scenario and two real \gls{iot} navigation and networking applications, showing that its performance is close to the full-knowledge \gls{tkf}'s.

The rest of this paper is organized as follows: in Sec.~\ref{sec:tkf}, the Tobit model and standard \gls{tkf} are presented. In Sec.~\ref{sec:atkf}, the \gls{atkf} is derived using the noise estimator in~\cite{gao2015adaptive}. The simulation results are shown in Sec.~\ref{sec:res}, and Sec.~\ref{sec:concl} concludes the paper and lists some possible avenues of future work.

\section{Related work}

Kalman filtering is one of the most used tools in \gls{iot} tracking applications, with examples in very different domain. Smart Grid management can be helped by the use of \glspl{kf} to track the voltage in a microgrid remotely even with noisy communication channels~\cite{rana2015kalman}, and in~\cite{nguyen2017effective}, a \gls{kf} is used to predict the amount of energy that the wireless nodes will harvest in the near future, guiding routing decisions. It can even be used to track traffic jams~\cite{gong2013research} or health-related cycling metrics and statistics~\cite{zhao2017real}. Kalman filtering can also aid distributed sensing, to improve the accuracy of the tracked variable or to reduce useless communications~\cite{huang2019epkf} by tracking the channel and the usefulness of the transmission.

Localization and ranging are two of the applications in which \glspl{kf} are used most often: while several methods to gauge the position and distance of a node from radio transmissions exist, from time of arrival estimation to \gls{rss}-based ranging, they are often inaccurate in complex propagation environments. \glspl{kf} have been used to reduce the noise and improve the estimate accuracy using \gls{ble}~\cite{ozer2016improving} and \gls{lora}~\cite{bakkali2017kalman} signals, and a node equipped with multiple wireless interfaces can use a \gls{kf} for sensor fusion to improve the overall positioning accuracy. The use of \gls{vlc} signals in localization has recently given some good results, using \glspl{kf} to reduce the measurement error~\cite{zhuang2019low}.

The \gls{kf} is a powerful tool, but its linearity requirement is restrictive; several approaches have been tried to deal with nonlinearities in the system and noise.
The first approach to the Tobit censoring issue was to consider censored measurements as missing, and adapt the \gls{kf} to deal with intermittent measurements~\cite{sinopoli2004kalman}. However, the performance degrades significantly when the state of the system is close to the censoring region and the censoring probability is high.
The \gls{tkf}~\cite{allik2015tobit} is a recently developed tool that can achieve good performance in Type I censored systems. It calculates the expected value of the measurements after taking the censoring probability into account and uses it to find a modified innovation value. The filter assumes that the state error is small to get a recursive formulation: the diagonality of the noise covariance matrices is assumed to simplify the notation, but not strictly required. The \gls{tkf} has been shown to outperform even nonlinear filters such as the \gls{ekf} and \gls{ukf}, with better estimates of the state uncertainty and smoother transitions from censoring to non-censoring~\cite{allik2015nonlinear}. A similar adaptation has been proposed for particle filters~\cite{allik2017particle}, another common filtering tool. Another work~\cite{loumponias2018adaptive} uses a modified \gls{tkf} to adapt to changes in the doubly censored case.

Another well-known issue of the traditional \gls{kf} is its requirement of full \emph{a priori} knowledge of the measurement and process noise covariances: in many applications, these covariances are hard to estimate in advance or even time-varying, and the design of an efficient \gls{akf} has been the subject of considerable research interest over the years, from the first works in the 1970s~\cite{mehra1970identification} to more recent approaches. The \gls{als} method~\cite{rajamani2009estimation} uses the autocovariance of the innovation signal to estimate the noise covariance matrices. It works in the general case of non-diagonal covariance matrices, and results in the optimal estimate, but it requires convergence of the Kalman gain before it can operate. As such, it is poorly equipped to deal with time-varying noise statistics or complex scenarios. The time-varying noise issue was solved by the unbiased one-step estimator in~\cite{gao2015adaptive}: while not fully exploiting the historical information, this estimator can react to shifts in the noise and deal with an unstable Kalman gain. Another recent work~\cite{zanni2016prediction} develops an \gls{arima} approach to the estimation of the process noise covariance with fast convergence when the measurement noise statistics are known.

The case of Tobit Kalman filtering with unkown parameters has been covered by a few recent works, investigating different aspects of the issue.
A modified \gls{tkf} presented in~\cite{geng2019state} can also deal with non-Gaussian L{\'e}vy and time-correlated measurement noise. In this case, the noise is transformed into a Gaussian noise with unknown variance, which is estimated by a recursive process, using the \emph{a priori} knowledge of the process noise covariance and the measurement noise time correlation.
Finally, the case of uncertainty in the system model is examined in~\cite{han2018improved}: the authors introduce some randomness in the dynamic system, using stochastic update and measurement matrices with known statistics and known noise covariance matrices.

\section{The Tobit Kalman Filter}\label{sec:tkf}
We consider a dynamic linear system with Tobit Type I censoring:
\begin{align}
 \mathbf{x}_{k+1}&=\mathbf{A}_k\mathbf{x}_k+\mathbf{w}_k\label{eq:newx}\\
 \mathbf{y}_{k}^*&=\mathbf{C}_k\mathbf{x}_k+\mathbf{v}_k\\
 \mathbf{y}_k&=\max(\bm{\tau}_k,\mathbf{y}_k^*)
 &,
\end{align}
where $\mathbf{x}_k\in\mathbb{R}^n$ is a hidden state vector, $\mathbf{y}_k\in\mathbb{R}^m$ is the measurement output vector, $\mathbf{A}_k\in\mathbb{R}^{n\times n}$ is the non-singular update matrix and $\mathbf{C}_k\in\mathbb{R}^{m\times m}$ is the measurement matrix. The two components $\mathbf{w}_k$ and $\mathbf{v}_k$ are multivariate Gaussian random vectors with zero mean and covariance matrices $\mathbf{Q}_k\in\mathbb{R}^{n\times n}$ and $\mathbf{R}_k\in\mathbb{R}^{m\times m}$, respectively. The standard \gls{kf} is the optimal estimator of the hidden state as long as there is no censoring, but the Tobit nonlinearity makes it suboptimal if some measurements are censored.

We now introduce the notation used in the following: given a random variable $X$, its expected value is denoted by $\mathbb{E}[X]$ and its variance is $\text{Var}[X]$. The conditional expectation of $X$ given the value of $Y$ is denoted by $\mathbb{E}[X|Y]$. Vectors like $\mathbf{x}$ are written in bold, while matrices like $\mathbf{A}$ are written in bold and indicated with capital letters. The hat symbol indicates that the value is an estimate: $\hat{x}$ is an estimate of $x$. We refer to the univariate normal \gls{pdf} as $\phi(\cdot)$, and to the normal \gls{cdf} as $\Phi(\cdot)$.
To simplify the notation in the following steps, we also define the elements of the vector $\bm{\eta}_k$ as:
\begin{align}
\eta_k(i)&= \frac{C_k\hat{x}_{k|k-1}(i)-\tau_k(i)}{\sigma_k(i)}
\end{align}
We also define the inverse Mills ratio, i.e., $\mathbb{E}[X|X>\alpha]$, which we denote by $\lambda(\alpha)$ for the univariate normal case:
\begin{align}
  \lambda(\alpha)&=\frac{\phi(\alpha)}{1-\Phi(\alpha)}.
\end{align}
Its variance equivalent $\mathbb{E}[X^2|X>\alpha]$ is denoted by $\eth(\alpha)$:
\begin{align}
  \eth(\alpha)&=\lambda(\alpha)\left(\lambda(\alpha) - \alpha\right).
\end{align}

Using the notation defined above, we now recall the derivation of the \gls{tkf} from~\cite{allik2015tobit}. The first two moments of $y_k(i)$ are given by:
\begin{align}
&\begin{aligned}
\mathbb{E}[y_k(i)|\mathbf{x}_k,\sigma_k(i)]=&(1-\Phi(\eta_k(i)))\tau_k(i)+\\
&\Phi(\eta_k(i))\left(C_kx_k(i)+\sigma_k(i)\lambda(-\eta_k(i))\right)\label{eq:meany}
\end{aligned}\\
&\text{Var}[y_k(i)|\mathbf{x}_k,\sigma_k(i)]=\sigma_k^2(i)\left[1-\eth\left(\eta_k(i)\right)\right],\label{eq:vary}
\end{align}
The Kalman error $\tilde{y}_k(i)$ is then given by:
\begin{align}
 \tilde{y}_k(i)&=y_k(i)-\mathbb{E}[y_k(i)|\mathbf{x}_{k|k-1},\sigma_k(i)].\label{eq:err}
\end{align}
Additionally, we use a diagonal $m\times m$ matrix of Bernoulli variables $\mathbf{P}_k$ to represent the censoring of the measurements. Its elements are given by:
\begin{align}
 p_k(i,j)&=I(y_k(i)>\tau_k(i))\delta_{i,j},
\end{align}
where $\delta_{i,j}$ is the Kronecker delta function and $I(\cdot)$ is the indicator function. The expected value of the censoring variable is:
\begin{align}
 \mathbb{E}[p_k(i,j)]&=\Phi\left(\eta_k(i)\right).
\end{align}
We now define the covariance matrix $\mathbf{R}_{\mathbf{\tilde{x}\tilde{y}},k}=\mathbb{E}\left[(\mathbf{x}_k-\mathbf{x}_{k-1})\mathbf{\tilde{y}}_k^T\right]$:
\begin{align}
 \mathbf{R}_{\mathbf{\tilde{x}\tilde{y}},k}&=\mathbf{\Psi}_{k|k-1}\mathbf{C}_k^T\mathbb{E}[\mathbf{P}_k],
\end{align}
where $\mathbf{\Psi}_{k|k-1}=\mathbb{E}[(\mathbf{x}_k-\mathbf{\hat{x}}_k)(\mathbf{x}_k-\mathbf{\hat{x}}_k)^T]$ is the predicted \emph{a priori} state error covariance.
We define the column vector $\mathbf{\tilde{v}}_k$ representing the bias introduced in the measurement noise by the censoring, whose elements are $\tilde{v}_k(i)=\sigma_k(i)\lambda\left(-\eta_k(i)\right)$. As in~\cite{allik2015tobit}, we define the Kalman error covariance matrix  $\mathbf{R}_{\mathbf{\tilde{y}\tilde{y}},k}=\mathbb{E}\left[\mathbf{\tilde{y}}_k\mathbf{\tilde{y}}_k^T\right]$:
\begin{align}
\begin{aligned}
\mathbf{R}_{\mathbf{\tilde{y}\tilde{y}},k}=&\mathbb{E}[\mathbf{P}_k]\mathbf{C}_k\mathbf{\Psi}_{k|k-1}\mathbf{C}_k^T\mathbb{E}[\mathbf{P}_k]+\\
&\mathbb{E}[\mathbf{P}_k(\mathbf{v}_k-\mathbf{\tilde{v}}_k)(\mathbf{v}_k-\mathbf{\tilde{v}}_k)^T\mathbf{P}_k].\label{eq:Ryy}
\end{aligned}
\end{align}
The second term of the sum corresponds to the measurement covariance matrix $\mathbf{V}_k$, which is equal to:
\begin{align}
 \mathbf{V}_k&=\text{diag}\begin{bmatrix}
                           \text{Var}[y_k(1)|\hat{x}_{k|k-1}(1),\sigma_k(1)]\\
                           \text{Var}[y_k(2)|\hat{x}_{k|k-1}(2),\sigma_k(2)]\\
                           \vdots\\
                           \text{Var}[y_k(m)|\hat{x}_{k|k-1}(m),\sigma_k(m)]
                          \end{bmatrix}
.
\end{align}
The derivation in~\cite{allik2015tobit} assumes that $\mathbf{Q}_k$ and $\mathbf{R}_k$ are diagonal to simplify the notation. The derivation of the theoretical covariance matrix in the general case can be adapted from the characterization of the truncated multivariate normal in~\cite{rosenbaum1961moments}. 
The Kalman gain is calculated as:
\begin{align}
 \mathbf{K}_j=\mathbf{R}_{\mathbf{\tilde{x}\tilde{y}},j}\mathbf{R}_{\mathbf{\tilde{y}\tilde{y}},j}^{-1}.
\end{align}
The \gls{tkf} is then given by:
\begin{align}
\mathbf{\hat{x}}_{k|k-1}&=\mathbf{A}_{k-1}\mathbf{\hat{x}}_{k-1|k-1}\label{eq:tkf_update}\\ 
\mathbf{\Psi}_{k|k-1}&=\mathbf{A}_{k-1}\mathbf{\Psi}_{k-1|k-1}\mathbf{A}_{k-1}^T+\mathbf{Q}_{k-1}\\
\mathbf{\hat{x}}_{k|k}&=\mathbf{\hat{x}}_{k|k-1}+\mathbf{R}_{\mathbf{\tilde{x}\tilde{y}},k}\mathbf{R}_{\mathbf{\tilde{y}\tilde{y}},k}^{-1}\mathbf{\tilde{y}}_k\label{eq:xupdate}\\ 
\mathbf{\Psi}_{k|k}&=\left(\mathbf{I}_{m\times m}-\mathbb{E}[\mathbf{P}_k])\mathbf{R}_{\mathbf{\tilde{x}\tilde{y}},k}\mathbf{R}_{\mathbf{\tilde{y}\tilde{y}},k}^{-1}\mathbf{C}_k\right)\mathbf{\Psi}_{k|k-1}\label{eq:tkf_p}
\end{align}

\section{A Posteriori Noise Covariance Estimation}\label{sec:atkf}
In this section, we derive the noise covariance estimator for the \gls{tkf}. In a system with diagonal noise covariance matrices, the \emph{a posteriori} density function of the noise covariance of a standard Kalman filter can be estimated using the \gls{map} coupling form~\cite{sage1969adaptive}:
\begin{align}
 \mathbf{\hat{Q}}_k&=\frac{1}{k}\sum_{\ell=1}^k\left[\mathbf{\hat{x}}_{k|\ell}-\mathbf{A}_{\ell-1}\mathbf{\hat{x}}_{k|\ell-1}\right]\left[\mathbf{\hat{x}}_{k|\ell}-\mathbf{A}_{\ell-1}\mathbf{\hat{x}}_{k|\ell-1}\right]^T\label{eq:q_est_opt}\\ 
 \mathbf{\hat{R}}_k&=\frac{1}{k}\sum_{\ell=1}^k\left[\mathbf{y}_{\ell}-\mathbf{C_\ell}\mathbf{\hat{x}}_{k|\ell}\right]\left[\mathbf{y}_{\ell}-\mathbf{C_\ell}\mathbf{\hat{x}}_{k|\ell}\right]^T.\label{eq:r_est_opt}
\end{align}
Even in the standard scenario with no censoring, the multistep terms $\mathbf{\hat{x}}_{k|\ell-1}$ and $\mathbf{\hat{x}}_{k|\ell}$ make the optimal formulation above impractical, as it cannot be described in recursive form. A practical one-step approximation is presented in~\cite{gao2015adaptive}. In the \gls{tkf} case, the derivation for the practical unbiased estimator for $\mathbf{\hat{Q}}_k$ follows from~\eqref{eq:newx}. The process noise covariance matrix is still $\mathbf{Q}=\mathbb{E}\left[(\mathbf{x}_k-\mathbf{A}_{k-1}\mathbf{x}_{k-1})(\mathbf{x}_k-\mathbf{A}_{k-1}\mathbf{x}_{k-1})^T\right]$, and if we assume that the \gls{atkf} state estimate is close enough to the real state, i.e., $\mathbf{\hat{x}}_{k|k}\simeq\mathbf{x}_k$, an assumption that the standard \gls{tkf} also requires, we get:
\begin{align}
\mathbf{Q}&=\mathbb{E}\left[(\mathbf{x}_k-\mathbf{A}_{k-1}\mathbf{x}_{k-1})(\mathbf{x}_k-\mathbf{A}_{k-1}\mathbf{x}_{k-1})^T\right]\\
\mathbf{Q}&\simeq\mathbb{E}\left[(\mathbf{\hat{x}}_{k|k}-\mathbf{A}_{k-1}\mathbf{\hat{x}}_{k-1|k-1})(\mathbf{\hat{x}}_{k|k}-\mathbf{A}_{k-1}\mathbf{\hat{x}}_{k-1|k-1})^T\right]\\
\mathbf{Q}&\simeq\mathbb{E}\left[(\mathbf{\hat{x}}_{k|k}-\mathbf{\hat{x}}_{k|k-1})(\mathbf{\hat{x}}_{k|k}-\mathbf{\hat{x}}_{k|k-1})^T\right]\\
\mathbf{Q}&\simeq\mathbb{E}\left[(\mathbf{K}_k\mathbf{\tilde{y}}_k)(\mathbf{K}_k\mathbf{\tilde{y}}_k)^T\right].
\end{align}
We can now estimate the value of $\mathbf{Q}$ from the available samples:
\begin{align}
 \mathbf{\hat{Q}}_k&=\frac{1}{k}\sum_{j=1}^k \mathbf{K}_j\mathbf{\tilde{y}}_j\mathbf{\tilde{y}}_j^T\mathbf{K}_j^T.\label{eq:q_biased}
\end{align}
The expected value of the process noise covariance is:
\begin{align}
 \mathbb{E}\left[\mathbf{\hat{Q}}_k\right]&=\mathbb{E}\left[\frac{1}{k}\sum_{j=1}^k\mathbf{K}_j\mathbf{\tilde{y}}_j\mathbf{\tilde{y}}_j^T\mathbf{K}_j^T\right]\\
 \mathbb{E}\left[\mathbf{\hat{Q}}_k\right]&=\frac{1}{k}\mathbb{E}\left[\sum_{j=1}^k\mathbb{E}[\mathbf{P}_j]\mathbf{K}_j\mathbf{C}_j\bm{\Psi}_{j|j-1}\right]\label{eq:nextstep}\\
 \mathbb{E}\left[\mathbf{\hat{Q}}_k\right]&=\mathbb{E}\left[\frac{1}{k}\sum_{j=1}^k\bm{\Psi}_{j|j-1}-\bm{\Psi}_{j|j}\right]\label{eq:covmean}
\end{align}
The derivation of~\eqref{eq:nextstep} follows from the fact that $\mathbb{E}\left[\mathbf{\tilde{y}}_j\mathbf{\tilde{y}}_j^T\right]=\mathbf{R}_{\mathbf{\tilde{y}\tilde{y}},j}$. In the following step, we use~\eqref{eq:tkf_p} to remove the Kalman gain from the equation.
We can now subtract the expected value in~\eqref{eq:covmean} from~\eqref{eq:q_biased} to write the unbiased estimator for the process noise covariance:
\begin{align}
  \mathbf{\hat{Q}}_k&=\frac{1}{k}\sum_{\ell=1}^k\mathbf{K}_j\mathbf{\tilde{y}}_j\mathbf{\tilde{y}}_j^T\mathbf{K}_j^T+\mathbf{\Psi}_{j|j}-\mathbf{A}\mathbf{\Psi}_{j-1|j-1}\mathbf{A}^T\label{eq:q_est_sub}.
\end{align}

The estimation noise covariance matrix $\mathbf{R}_k$ does not have a linear estimator, since the Kalman error covariance matrix $\mathbf{R}_{\mathbf{\tilde{y}\tilde{y}},k}$ is given by~\eqref{eq:Ryy}, which contains $\mathbf{V}_k$ instead of $\mathbf{R}_k$. We get the unbiased estimator for $\mathbf{\hat{V}}_k$ by substituting the \gls{tkf} equations into the one-step version of~\eqref{eq:r_est_opt}:
\begin{align}
   \mathbf{\hat{V}}_k&=\frac{1}{k}\sum_{\ell=1}^k\left(\mathbf{I}-\mathbf{C}_j\mathbf{K}_j\right)\left(\mathbf{\tilde{y}}_j\mathbf{\tilde{y}}_j^T\left(\mathbf{I}-\mathbf{C}_j\mathbf{K}_j\right)+\mathbf{C}_j\mathbf{R}_{\mathbf{\tilde{x}\tilde{y}},j}\right).\label{eq:r_est_sub}
\end{align}
The derivation of~\eqref{eq:r_est_sub} is the same as in~\cite{gao2015adaptive}, using the \gls{tkf} modified equations instead of the standard \gls{kf}.
In order to estimate $\mathbf{\hat{R}}_k$, we can now simply invert~\eqref{eq:vary}:
\begin{align}
 \hat{R}_k(i,i)&=\frac{\hat{V}_k(i,i)}{1-\eth(\hat{\eta}_k(i))}\label{eq:inverse}\\
 \hat{\eta}_k(i)&=\frac{C_k\hat{x}_{k|k-1}(i)-\tau_k(i)}{\hat{\sigma}_{k-1}(i)}.
\end{align}
The resulting heuristic estimator is not unbiased, since the function is non-linear, but corresponds to the one-step \gls{mle} for the censored Gaussian distribution, as derived by Gupta~\cite{gupta1952estimation}.

The \gls{tkf} converges to the standard \gls{kf} when the censoring region is far from the state value. The noise estimator also converges to the unbiased noise estimator used in the \gls{akf}. After substituting the terms in~\eqref{eq:q_est_sub}, the estimator is the one in~\cite{gao2015adaptive}. The same goes for the measurement noise estimator, since $\lim_{\bm{\eta}_k\rightarrow-\infty}\mathbf{\hat{R}}_k=\mathbf{\hat{V}}_k$.

Note also that the noise estimation is computationally simple, as its only additional load with respect to the \gls{akf} is the calculation of the inverse functions in~\eqref{eq:inverse}.Therefore, the \gls{atkf} can operate while requiring the extra computation of $m$ normal \glspl{pdf} and \glspl{cdf} at each iteration.

\subsection{Time-varying noise estimation}
In order to deal with time-varying noise, the estimation of the noise covariances needs to be performed over a limited window in time. This can be done with a simple lowpass filter, so that the new estimate of the covariance is the linear combination of the old estimate and the latest sample, weighted by a factor $\Gamma_k$:
\begin{align}
  \Gamma_k&=\frac{1-\gamma}{1-\gamma^k}\label{eq:gamma},
\end{align}
where $\gamma\in[0,1)$ is a fading factor. Lower values of the fading factor correspond to a higher weight to new samples, and setting $\gamma=0$ is equivalent to only considering the latest sample.
However, the covariance samples are sensitive to noise outliers, and the estimator might even diverge. For this reason, the estimator can use an \gls{ice} that averages the covariance samples over a rectangular sliding window to guarantee that noise covariance matrices are semidefinite positive~\cite{jwo2008adaptive}:
\begin{align}
 \bm{\xi}_k=\frac{1}{N}\sum_{j=k-N+1}^{k}\left[\mathbf{\tilde{y}}_j\mathbf{\tilde{y}}_j^T\right],\label{eq:ice}
\end{align}
where $N$ is the size of the sliding window. The \gls{ice} is another lowpass filter, and its length determines the reactiveness of the estimator. If the noise statistics are fast-varying, shorter windows are recommended. It is now possible to rewrite the estimator recursive equations, considering~\eqref{eq:gamma} and~\eqref{eq:ice}:
\begin{align}
 &\begin{aligned}
   \mathbf{\hat{Q}}_k=(1-&\Gamma_k)\mathbf{\hat{Q}}_{k-1}+\Gamma_k\cdot\\ &\left(\mathbf{K}_k\bm{\xi}_k\mathbf{K}_k^T+\mathbf{\Psi}_{k|k}-\mathbf{A}_k\mathbf{\Psi}_{k-1|k-1}\mathbf{A}_k^T\right)\label{eq:qsample}\\
 \end{aligned}\\
 &\mathbf{\hat{V}}_k=\left(\mathbf{I}-\mathbf{C}_k\mathbf{K}_k\right)\left(\mathbf{K}_k\bm{\xi}_k\left(\mathbf{I}-\mathbf{C}_k\mathbf{K}_k\right)+\mathbf{C}_k\mathbf{R}_{\mathbf{\tilde{x}\tilde{y}},k}\right)\\
 &\mathbf{\hat{R}}_k=(1-\Gamma_k)\mathbf{\hat{R}}_{k-1}+\Gamma_k\left(\mathbf{I}-\text{diag}\left(\eth\left(\bm{\hat{\eta}}_k\right)\right)\right)^{-1}\hat{\mathbf{V}}_k\label{eq:fullr}
\end{align}
The full \gls{atkf} is given by the full estimator in~\eqref{eq:qsample}-\eqref{eq:fullr}, along with the update formulas in~\eqref{eq:tkf_update}-\eqref{eq:tkf_p}.

\section{Performance evaluation}\label{sec:res}

In this section, we present the results from two simulations, whose parameters are taken from~\cite{allik2015tobit} and~\cite{huang2017structure}. The first scenario is simple and shows the limits of the Kalman approach in censored scenarios, and how the Tobit version of the filter can overcome them. In the second, a positioning application is extended to include the censored case, in which the position information is not available outside a limited region. We also included the original two-dimensional scenario from~\cite{allik2015tobit} in Appendix~\ref{sec:ballistic} to provide a more complete comparison, even though it is not related to \gls{iot} scenarios. The \gls{atkf} will be compared with three other filters: the standard \gls{kf} and the \gls{tkf} with the correct values of $\mathbf{Q}_k$ and $\mathbf{R}_k$, and the \gls{akf} from~\cite{gao2015adaptive}. In this way, we will show that the \gls{atkf} achieves almost the same performance as the \gls{tkf} while estimating the noise covariances online. We set $\gamma=0.33$ and $N=30$ in both examples.

\subsection{Constant value estimation}
\begin{figure}[!t]
 \centering
  \includegraphics[trim=0 0.1cm 0 0]{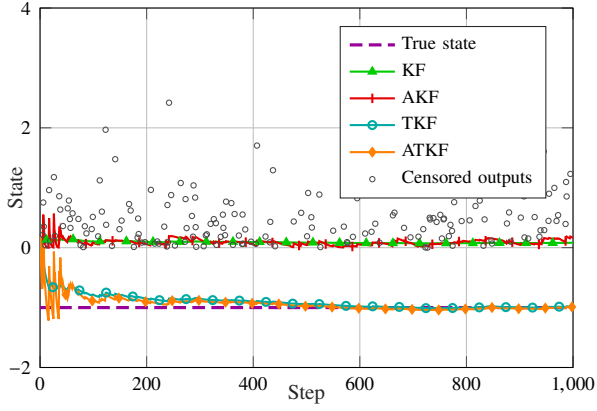}
 \caption{1D example: estimation of a constant value one standard deviation of the measurement noise below the censoring limit.}
\label{fig:evo_1d}
\end{figure}

In the first example, we will estimate a 1D constant value below the censoring limit. In this case, we have $x=-1$, $Q=0$, $R=1$, and $\tau=0$. This means that any negative measurement value will be censored, and the state is one standard deviation below the limit. As Fig.~\ref{fig:evo_1d} shows, most of the samples are censored. The initial conditions for the filters are $\hat{x}_0=5$ and $\hat{\Psi}_0=25$, and the two adaptive filters are initialized with $\hat{Q}_0=1$ and $\hat{R}_0=1$. Fig.~\ref{fig:evo_1d} shows that the \gls{kf} and \gls{akf} have a similar performance, staying above the censoring limit and having a significant error, while the \gls{tkf} estimates the state correctly after relatively few steps. The \gls{atkf} has a noisier evolution for the first 50 steps, but quickly converges to the true value of $x$.

\begin{figure}[!t]
 \centering
  \includegraphics[trim=0 0.1cm 0 0]{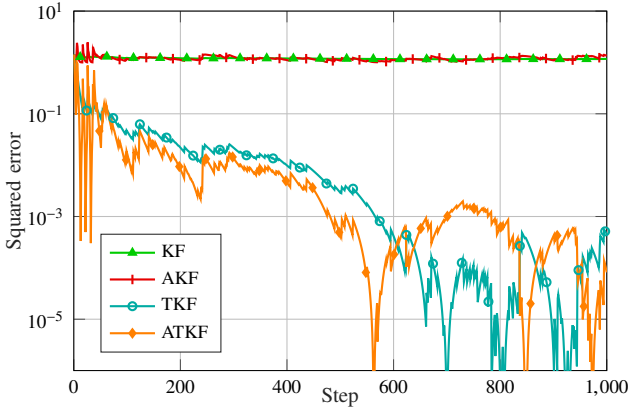}
 \caption{1D example: squared error of the filters.}
\label{fig:err_1d}
\end{figure}

Fig.~\ref{fig:err_1d} shows the evolution of the squared error for the different filters. The error is larger than 1 for the \gls{kf} and \gls{akf}, whose estimates are always above the censoring threshold, while the error of the \gls{tkf} and \gls{atkf} soon reaches very small values. In this case, the \gls{atkf} even performs slightly better than the \gls{tkf} after the initial phase, with a lower~\gls{mse}, but both errors are negligible.

\subsection{VLC positioning}
A classic \gls{iot} application is indoor positioning: mobile robots and simple sensor nodes can combine their own internal sensors and the parameters of a wireless signal received from a known base station through sensor fusion, using the \gls{rss}~\cite{zanella2016best} or \gls{aoa}~\cite{niculescu2004positioning} to improve their localization accuracy. Over the past few years, \gls{vlc} measurements have become a viable alternative to classic WiFi or \gls{ble}-based positioning methods~\cite{rahaim2012state}, and Kalman filters have already been used to improve tracking in this kind of application~\cite{vatansever2017visible}. 

The third example we show uses \gls{vlc} to estimate a mobile robot's position and an internal gyroscope to measure its heading. The dynamic system we use is the same as in~\cite{nguyen2014improvement}, with the same parameters, which are reported in Table~\ref{tab:vlcparams}. We consider a timestep $T$, and the state $\mathbf{x}_k$ of the robot at step $k$ is denoted by its position $(x_{k,1}, x_{k,2})$ and its heading $\theta_k$. We assume that it has two independent wheels, with a radius $R_w$ and placed at a distance $d_w$. The robot can maintain a speed $v$, and its turning rate is $\Delta\theta$. The motion update equation is given by:
\begin{align}
 \begin{bmatrix}x_{k+1,1}\\x_{k+1,2}\\ \theta_{k+1}\end{bmatrix}&=\begin{bmatrix}x_{k,1}\\x_{k,2}\\ \theta_{k}\end{bmatrix}+\begin{bmatrix}vT\cos\left(\theta_k+\frac{\Delta\theta}{2}\right)\\vT\sin\left(\theta_k+\frac{\Delta\theta}{2}\right)\\ \Delta\theta_{k}\end{bmatrix}+\mathbf{w}_k,\label{eq:vlc_update}
\end{align}
where $\mathbf{w}_k\sim \mathcal{N}(0,\mathbf{Q})$ is the Gaussian process noise. The Kalman update matrix $\mathbf{A}_{k+1}$ is the Jacobian matrix of the update function, containing its partial derivatives to the state $\mathbf{x}_k$:
\begin{align}
 \mathbf{A}_{k+1}&=\begin{bmatrix}1 & 0 & -vT\sin\left(\theta_k+\frac{\Delta\theta}{2}\right)\\ 0 & 1 & vT\cos\left(\theta_k+\frac{\Delta\theta}{2}\right)\\ 0 & 0 & 1\end{bmatrix}.
\end{align}
The measurement matrix $\mathbf{C}=\mathbf{I}_{3\times3}$ gives the following measurement function:
\begin{align}
 \mathbf{y}_k&=\mathbf{x}_k+\mathbf{v}_k,
\end{align}
where $\mathbf{v}_k\sim \mathcal{N}(0,\mathbf{R})$ is the Gaussian measurement noise.

We can now define the covariance matrices of the two noise components:
\begin{align}
 \mathbf{Q}_k&=vTk_w\mathbf{W}\mathbf{W}^T,
\end{align}
where $\mathbf{W}_k$ is the Jacobian matrix of the update function given in~\eqref{eq:vlc_update} with respect to $\mathbf{w}_k$:
\begin{align}
 W_{k,i,j}&=\frac{\partial x_{k,i}}{\partial w_{k,j}}.
\end{align}
In this case, $\mathbf{Q}_k$ is not symmetrical, so the \gls{atkf} will have an additional source of errors. The measurement noise covariance matrix $\mathbf{R}_k$ is defined as:
\begin{align}
 \mathbf{R_k}=\begin{bmatrix}\sigma_{\text{VLC}}^2 & 0 & 0\\ 0 & \sigma_{\text{VLC}}^2 & 0\\ 0 & 0 & \sigma_{\text{gyro}}^2\end{bmatrix},
\end{align}
where $\sigma_{\text{VLC}}^2$ is the \gls{vlc} positioning error variance and $\sigma_{\text{gyro}}^2$ is the gyroscopic attitude measurement error variance.

\begin{table}[b!]
	\centering
	\scriptsize
	\setstretch{1.2}
		\caption{VLC positioning parameters.}
	\label{tab:vlcparams}
	\begin{tabular}{ccc}
		\toprule
		Parameter & Value & Description\\ \midrule
		$R_w$ & 5 cm & Robot wheel radius\\
		$d_w$ & 30 cm & Robot wheel distance\\
		$T$ & 0.05 s & Filter timestep\\
		$k_w$ & 0.0003 & Wheel-floor interaction parameter\\
		$\sigma_{\text{VLC}}$ & 0.06 cm & VLC positioning error\\
		$\sigma_{\text{gyro}}$ & 3$^\circ$ & Gyroscopic attitude measurement error\\
		$\theta_0$ & 45$^\circ$ & Initial heading \\
		$\Delta\theta$ & 0$^\circ$/s & Robot turning rate\\
		$v$ & 1 m/s & Robot speed\\
		\bottomrule
	\end{tabular}
\end{table}

\begin{figure}[!t]
 \centering
  \includegraphics[trim=0 0.1cm 0 0]{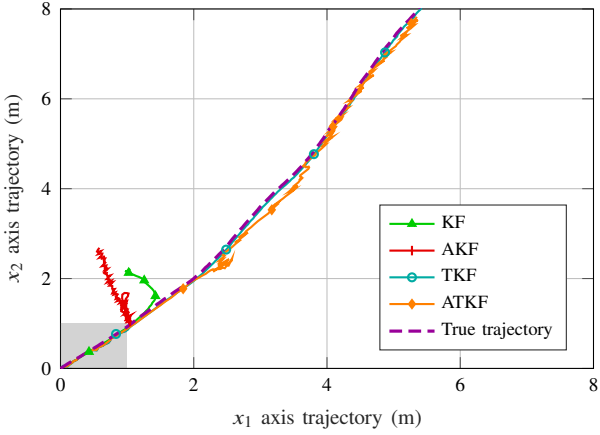}
 \caption{VLC positioning: estimated trajectories.}
\label{fig:vlc_traj}
\end{figure}

\begin{figure}[!t]
 \centering
  \includegraphics[trim=0 0.1cm 0 0]{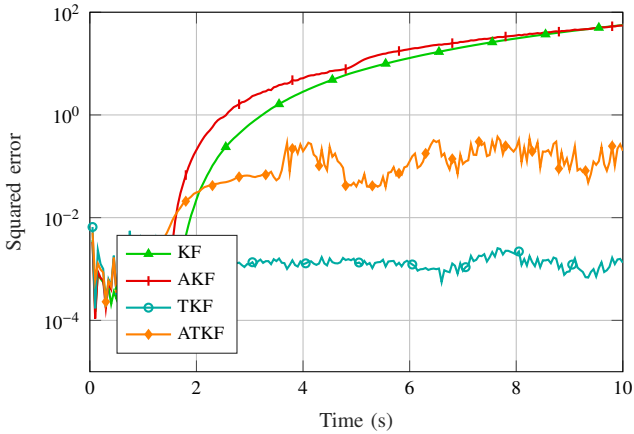}
 \caption{VLC positioning: squared error of the filters.}
\label{fig:err_vlc}
\end{figure}

In our system, we consider a single \gls{vlc} transmitter, whose signal can only be sensed underneath a roof cover: if the robot moves outside the covered area, sunlight will interfere with the \gls{vlc} system~\cite{nlom2017evaluation}. making the localization measurement unusable. The roof is considered as a square of 1 m, and the \gls{vlc} transmitter is placed at one of its vertices. The effect of the interference is shown in Fig.~\ref{fig:vlc_traj}: measurements are only available in the shaded area, and the values outside are censored. The plot clearly shows that the \gls{tkf} and \gls{atkf} are able to follow the robot's path, while the standard filters
Even if the \gls{tkf} and \gls{atkf} seem to follow the trajectory correctly, Fig.~\ref{fig:err_vlc} shows that the \gls{atkf}'s timing is off, resulting in a 0.5 m offset in its position estimate while the \gls{tkf} remains very precise. However, the performance of the filter is still very good considering the extremely limited amount of information available to it.

\section{Conclusion}\label{sec:concl}
This work presents a recursive noise covariance estimation method for the \gls{tkf}, showing a small performance loss with respect to the \gls{tkf} with full \emph{a priori} knowledge. The \gls{atkf} is inspired by the approach from~\cite{gao2015adaptive}, it does not add significant computational complexity to a standard \gls{tkf}, and it also converges to the standard \gls{akf} when the state is far from the censoring region. The estimate of the measurement noise is biased because of the non-linearity of the relation between innovation variance and measurement noise variance, but it is optimal if we consider a one-step memory. The estimation method is also robust to time-varying noise statistics and censoring thresholds.
The \gls{atkf} is tested in a simple example and an \gls{iot} positioning application, and shown to be a versatile and powerful tool to improve the estimation and tracking of variables with censored sensor data, which often occur in this networking paradigm.

Future work on the subject includes the adaptation to the Tobit Type I model of other \gls{akf} methods such as \gls{als}, which might outperform the recursive approach in some scenarios. The implementation of the \gls{atkf} in actual \gls{iot} devices is also an interesting subject of research. Finally, the extension to non-diagonal noise covariance matrices will be considered in an extension of this work.

\appendix
\section{Magnetometer attitude estimation}\label{sec:ballistic}

In this appendix, we provide another comparison with the \gls{tkf}, using the scenario from the original work~\cite{allik2015tobit}, which models ballistic roll using censored magnetometer measurements~\cite{allik2013ballistic}. This kind of system is often used in \glspl{uav} to estimate the attitude~\cite{huang2017structure}, and Kalman-based tracking has already been proposed in the literature~\cite{de2011uav}. The dynamic system in this example can be easily adapted to a fixed-wing \gls{uav} performing banking maneuvers~\cite{euston2008complementary}, and even to a ground vehicle making sharp turns~\cite{garcia2018design}. The measurement is one-dimensional ($m=1$), and it is derived from a 2D hidden state ($n=2$). In this case, we need to use an \gls{ekf}, as the equations of the system are nonlinear.
\begin{align}
 \mathbf{A}_k&=\alpha\begin{bmatrix}
                           \cos(\omega)\quad-\sin(\omega)\\
                           \sin(\omega)\quad\cos(\omega)
                          \end{bmatrix}\\
 \mathbf{C}_k&=\begin{bmatrix}
                           1\quad 0
                          \end{bmatrix} 
\end{align}
The oscillator frequency is $\omega=0.005\cdot2\pi$ and the gain is $\alpha=1$, with a sampling period $T=1$. The censoring threshold is still $\tau=0$.
In this case, we set $\mathbf{Q}_k=\mathbf{I}\cdot 0.0025$ and $\sigma_k^2=1$. The presence of a low process noise makes the system more irregular, and thus harder to estimate correctly when the measurements are censored. The initial conditions of the filters are $\mathbf{\hat{x}}_0=[5\ 0]^T$ and $\bm{\Psi}_0=\mathbf{I}$. As in the previous example, the adaptive filters are initialized with $\hat{\mathbf{Q}}_0=\mathbf{I}$ and $\hat{R}_0=1$.

\begin{figure}[!t]
 \centering
  \includegraphics[trim=0 0.1cm 0 0]{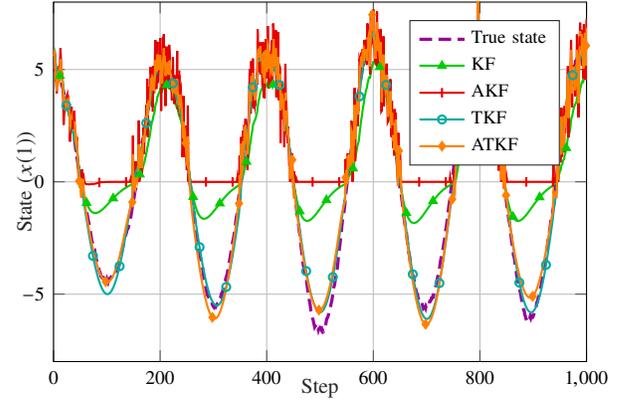}
 \caption{Attitude estimation: estimation of the observed state component.}
\label{fig:evo_2d1}
\end{figure}

\begin{figure}[!t]
 \centering
  \includegraphics[trim=0 0.1cm 0 0]{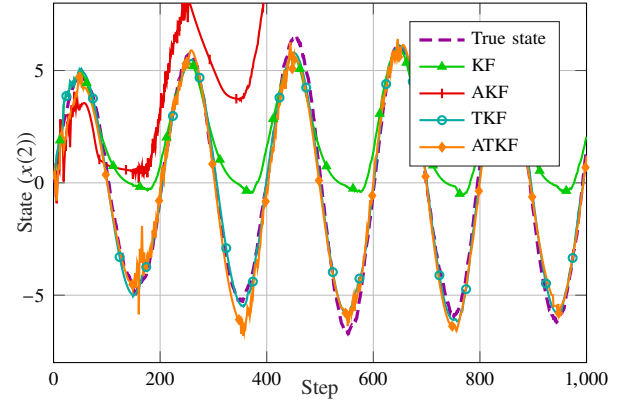}
 \caption{Attitude estimation: estimation of the hidden state component.}
\label{fig:evo_2d2}
\end{figure}

\begin{figure}[!t]
 \centering
  \includegraphics[trim=0 0.1cm 0 0]{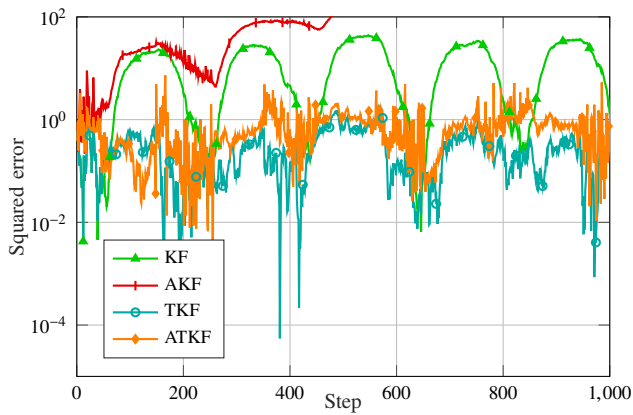}
 \caption{Attitude estimation: squared error of the filters.}
\label{fig:err_2d}
\end{figure}

Fig.~\ref{fig:evo_2d1} shows the evolution of the first state variable. When the measurements are not censored, the \gls{kf} is the only one that does not correctly follow the trend of the hidden state, even if the \gls{akf} is noisier than either the \gls{tkf} or the \gls{atkf}. However, the \gls{akf} never goes below 0 in its estimate, while the \gls{kf} manages to follow the state in its valleys, even as it strongly underestimates their magnitude. The \gls{tkf} and \gls{atkf} also make an error on the valley amplitude, probably because of the process noise, but the error is far lower.
Fig.~\ref{fig:evo_2d2} shows the evolution of the second state variable, which is not observed directly. In this case, the \gls{akf} quickly diverges, as it is unable to get a correct estimate of the noise covariances. The \gls{kf} does better, underestimating the valleys but correctly estimating the peaks. On the other hand, the \gls{tkf} and \gls{atkf} only make small mistakes on the amplitudes of the minima and maxima, correctly following the trend even when the measurement of the first state variable is censored and the second one is hidden.

Fig.~\ref{fig:err_2d} shows the evolution of the squared error in this example: in this case, the \gls{atkf} does slightly worse than the \gls{tkf}, with a \gls{mse} of 0.75, while the \gls{tkf} has 0.34. However, the error of the two filters is still lower than the measurement noise variance, even when the noise is censored. At the same time, the divergence of the \gls{akf} and the underestimation of the censored lobes by the standard \gls{kf} make them have a far higher error.


\bibliographystyle{IEEEtran}
\bibliography{bibliography}

\begin{thebibliography}{10}
\providecommand{\url}[1]{#1}
\csname url@samestyle\endcsname
\providecommand{\newblock}{\relax}
\providecommand{\bibinfo}[2]{#2}
\providecommand{\BIBentrySTDinterwordspacing}{\spaceskip=0pt\relax}
\providecommand{\BIBentryALTinterwordstretchfactor}{4}
\providecommand{\BIBentryALTinterwordspacing}{\spaceskip=\fontdimen2\font plus
\BIBentryALTinterwordstretchfactor\fontdimen3\font minus
  \fontdimen4\font\relax}
\providecommand{\BIBforeignlanguage}[2]{{%
\expandafter\ifx\csname l@#1\endcsname\relax
\typeout{** WARNING: IEEEtran.bst: No hyphenation pattern has been}%
\typeout{** loaded for the language `#1'. Using the pattern for}%
\typeout{** the default language instead.}%
\else
\language=\csname l@#1\endcsname
\fi
#2}}
\providecommand{\BIBdecl}{\relax}
\BIBdecl

\bibitem{gubbi2013internet}
J.~Gubbi, R.~Buyya, S.~Marusic, and M.~Palaniswami, ``{Internet of Things
  (IoT)}: A vision, architectural elements, and future directions,''
  \emph{Future Generation Computer Systems}, vol.~29, no.~7, pp. 1645--1660,
  Sep. 2013.

\bibitem{zhang2012energy}
L.~Zhang, J.~Liu, and H.~Jiang, ``Energy-efficient location tracking with
  smartphones for {IoT},'' in \emph{Sensors}.\hskip 1em plus 0.5em minus
  0.4em\relax IEEE, Oct. 2012, pp. 1--4.

\bibitem{gite2016context}
S.~Gite and H.~Agrawal, ``On context awareness for multisensor data fusion in
  {IoT},'' in \emph{2nd International Conference on Computer and Communication
  Technologies}.\hskip 1em plus 0.5em minus 0.4em\relax Springer, Jul. 2016,
  pp. 85--93.

\bibitem{kumar2013automatic}
D.~Kumar, S.~Rajasegarar, and M.~Palaniswami, ``Automatic sensor drift
  detection and correction using {Spatial Kriging and Kalman} filtering,'' in
  \emph{International Conference on Distributed Computing in Sensor
  Systems}.\hskip 1em plus 0.5em minus 0.4em\relax IEEE, May 2013, pp.
  183--190.

\bibitem{macagnano2014indoor}
D.~Macagnano, G.~Destino, and G.~Abreu, ``Indoor positioning: A key enabling
  technology for {IoT} applications,'' in \emph{World Forum on Internet of
  Things (WF-IoT)}.\hskip 1em plus 0.5em minus 0.4em\relax IEEE, Mar. 2014, pp.
  117--118.

\bibitem{jara2014mobile}
A.~J. Jara, P.~Lopez, D.~Fernandez, J.~F. Castillo, M.~A. Zamora, and A.~F.
  Skarmeta, ``Mobile digcovery: discovering and interacting with the world
  through the {Internet of Things},'' \emph{Personal and Ubiquitous Computing},
  vol.~18, no.~2, pp. 323--338, Feb. 2014.

\bibitem{perera2015energy}
C.~Perera, D.~S. Talagala, C.~H. Liu, and J.~C. Estrella, ``Energy-efficient
  location and activity-aware on-demand mobile distributed sensing platform for
  sensing as a service in {IoT} clouds,'' \emph{IEEE Transactions on
  Computational Social Systems}, vol.~2, no.~4, pp. 171--181, Dec. 2015.

\bibitem{lin2016enhanced}
K.~Lin, M.~Chen, J.~Deng, M.~M. Hassan, and G.~Fortino, ``Enhanced
  fingerprinting and trajectory prediction for {IoT} localization in smart
  buildings,'' \emph{IEEE Transactions on Automation Science and Engineering},
  vol.~13, no.~3, pp. 1294--1307, Apr. 2016.

\bibitem{bui2017survey}
N.~Bui, M.~Cesana, S.~A. Hosseini, Q.~Liao, I.~Malanchini, and J.~Widmer, ``A
  survey of anticipatory mobile networking: Context-based classification,
  prediction methodologies, and optimization techniques,'' \emph{IEEE
  Communications Surveys \& Tutorials}, vol.~19, no.~3, pp. 1790--1821, Apr.
  2017.

\bibitem{kalman1960new}
R.~E. Kalman, ``A new approach to linear filtering and prediction problems,''
  \emph{Journal of Basic Engineering}, vol.~82, no.~1, pp. 35--45, Mar. 1960.

\bibitem{yu2017precise}
N.~Yu, X.~Zhan, S.~Zhao, Y.~Wu, and R.~Feng, ``A precise dead reckoning
  algorithm based on {Bluetooth} and multiple sensors,'' \emph{IEEE Internet of
  Things Journal}, vol.~5, no.~1, pp. 336--351, 2017.

\bibitem{weng2006video}
S.-K. Weng, C.-M. Kuo, and S.-K. Tu, ``Video object tracking using adaptive
  {Kalman} filter,'' \emph{Journal of Visual Communication and Image
  Representation}, vol.~17, no.~6, pp. 1190--1208, Dec. 2006.

\bibitem{dellaert1997robust}
F.~Dellaert and C.~Thorpe, ``Robust car tracking using {Kalman} filtering and
  {Bayesian} templates,'' in \emph{Conference on Intelligent Transportation
  Systems (ITSC)}, vol.~1.\hskip 1em plus 0.5em minus 0.4em\relax IEEE, Oct.
  1997, pp. 72--83.

\bibitem{tobin1958estimation}
J.~Tobin, ``Estimation of relationships for limited dependent variables,''
  \emph{Econometrica: Journal of the Econometric Society}, vol.~26, no.~1, pp.
  24--36, Jan. 1958.

\bibitem{allik2015tobit}
B.~Allik, C.~Miller, M.~J. Piovoso, and R.~Zurakowski, ``The {Tobit Kalman}
  filter: an estimator for censored measurements,'' \emph{IEEE Transactions on
  Control Systems Technology}, vol.~24, no.~1, pp. 365--371, Jun. 2015.

\bibitem{gao2015adaptive}
W.~Gao, J.~Li, G.~Zhou, and Q.~Li, ``Adaptive {Kalman} filtering with recursive
  noise estimator for integrated {SINS/DVL} systems,'' \emph{The Journal of
  Navigation}, vol.~68, no.~1, pp. 142--161, Jan. 2015.

\bibitem{rana2015kalman}
M.~M. Rana and L.~Li, ``Kalman filter based microgrid state estimation using
  the {Internet of Things} communication network,'' in \emph{12th International
  Conference on Information Technology-New Generations}.\hskip 1em plus 0.5em
  minus 0.4em\relax IEEE, Apr. 2015, pp. 501--505.

\bibitem{nguyen2017effective}
T.~D. Nguyen, J.~Y. Khan, and D.~T. Ngo, ``An effective energy-harvesting-aware
  routing algorithm for {WSN-based IoT} applications,'' in \emph{International
  Conference on Communications (ICC)}.\hskip 1em plus 0.5em minus 0.4em\relax
  IEEE, May 2017, pp. 1--6.

\bibitem{gong2013research}
Y.-s. Gong and Y.~Zhang, ``Research of short-term traffic volume prediction
  based on {Kalman} filtering,'' in \emph{6th International Conference on
  Intelligent Networks and Intelligent Systems (ICINIS)}.\hskip 1em plus 0.5em
  minus 0.4em\relax IEEE, Nov. 2013, pp. 99--102.

\bibitem{zhao2017real}
Y.-X. Zhao, Y.-S. Su, and Y.-C. Chang, ``A real-time bicycle record system of
  ground conditions based on {Internet of Things},'' \emph{IEEE Access},
  vol.~5, pp. 17\,525--17\,533, Aug. 2017.

\bibitem{huang2019epkf}
Y.~Huang, W.~Yu, E.~Ding, and A.~Garcia-Ortiz, ``Epkf: Energy efficient
  communication schemes based on kalman filter for iot,'' \emph{IEEE Internet
  of Things Journal}, Feb. 2019.

\bibitem{ozer2016improving}
A.~Ozer and E.~John, ``Improving the accuracy of {Bluetooth Low Energy} indoor
  positioning system using {Kalman} filtering,'' in \emph{International
  Conference on Computational Science and Computational Intelligence
  (CSCI)}.\hskip 1em plus 0.5em minus 0.4em\relax IEEE, Dec. 2016, pp.
  180--185.

\bibitem{bakkali2017kalman}
W.~Bakkali, M.~Kieffer, M.~Lalam, and T.~Lestable, ``Kalman filter-based
  localization for {Internet of Things LoRaWAN} end points,'' in \emph{28th
  Annual International Symposium on Personal, Indoor, and Mobile Radio
  Communications (PIMRC)}.\hskip 1em plus 0.5em minus 0.4em\relax IEEE, Oct.
  2017, pp. 1--6.

\bibitem{zhuang2019low}
Y.~Zhuang, Q.~Wang, M.~Shi, P.~Cao, L.~Qi, and J.~Yang, ``Low-power
  centimeter-level localization for indoor mobile robots based on {Ensemble
  Kalman} smoother using {Received Signal Strength},'' \emph{IEEE Internet of
  Things Journal}, 2019.

\bibitem{sinopoli2004kalman}
B.~Sinopoli, L.~Schenato, M.~Franceschetti, K.~Poolla, M.~I. Jordan, and S.~S.
  Sastry, ``Kalman filtering with intermittent observations,'' \emph{IEEE
  Transactions on Automatic Control}, vol.~49, no.~9, pp. 1453--1464, Sep.
  2004.

\bibitem{allik2015nonlinear}
B.~Allik, C.~Miller, M.~J. Piovoso, and R.~Zurakowski, ``Nonlinear estimators
  for censored data: a comparison of the {EKF}, the {UKF} and the {Tobit
  Kalman} filter,'' in \emph{American Control Conference (ACC)}.\hskip 1em plus
  0.5em minus 0.4em\relax IEEE, Jul. 2015, pp. 5146--5151.

\bibitem{allik2017particle}
B.~Allik, ``Particle filter for target localization and tracking leveraging
  lack of measurement,'' in \emph{American Control Conference (ACC)}.\hskip 1em
  plus 0.5em minus 0.4em\relax IEEE, May 2017, pp. 1592--1597.

\bibitem{loumponias2018adaptive}
K.~Loumponias, A.~Dimou, N.~Vretos, and P.~Daras, ``Adaptive {Tobit
  Kalman}-based tracking,'' in \emph{14th International Conference on
  Signal-Image Technology \& Internet-Based Systems (SITIS)}.\hskip 1em plus
  0.5em minus 0.4em\relax IEEE, Jun. 2018, pp. 70--76.

\bibitem{mehra1970identification}
R.~Mehra, ``On the identification of variances and adaptive kalman filtering,''
  \emph{IEEE Transactions on Automatic Control}, vol.~15, no.~2, pp. 175--184,
  Apr. 1970.

\bibitem{rajamani2009estimation}
M.~R. Rajamani and J.~B. Rawlings, ``Estimation of the disturbance structure
  from data using semidefinite programming and optimal weighting,''
  \emph{Automatica}, vol.~45, no.~1, pp. 142--148, Jan. 2009.

\bibitem{zanni2016prediction}
L.~Zanni, J.-Y. Le~Boudec, R.~Cherkaoui, and M.~Paolone, ``A prediction-error
  covariance estimator for adaptive kalman filtering in step-varying processes:
  application to power-system state estimation,'' \emph{IEEE Transactions on
  Control Systems Technology}, vol.~25, no.~5, pp. 1683--1697, Dec. 2016.

\bibitem{geng2019state}
H.~Geng, Z.~Wang, Y.~Cheng, F.~E. Alsaadi, and A.~M. Dobaie, ``State estimation
  under non-{Gaussian L{\'e}vy} and time-correlated additive sensor noises: A
  modified {Tobit Kalman} filtering approach,'' \emph{Signal Processing}, vol.
  154, pp. 120--128, Jan. 2019.

\bibitem{han2018improved}
F.~Han, H.~Dong, Z.~Wang, G.~Li, and F.~E. Alsaadi, ``Improved {Tobit Kalman}
  filtering for systems with random parameters via conditional expectation,''
  \emph{Signal Processing}, vol. 147, pp. 35--45, Jun. 2018.

\bibitem{rosenbaum1961moments}
S.~Rosenbaum, ``Moments of a truncated bivariate normal distribution,''
  \emph{Journal of the Royal Statistical Society: Series B (Methodological)},
  vol.~23, no.~2, pp. 405--408, Jul. 1961.

\bibitem{sage1969adaptive}
A.~P. Sage and G.~W. Husa, ``Adaptive filtering with unknown prior
  statistics,'' in \emph{Joint Automatic Control Conference}, no.~7, Aug. 1969,
  pp. 760--769.

\bibitem{gupta1952estimation}
A.~Gupta, ``Estimation of the mean and standard deviation of a normal
  population from a censored sample,'' \emph{Biometrika}, vol.~39, no. 3/4, pp.
  260--273, Dec. 1952.

\bibitem{jwo2008adaptive}
D.-J. Jwo and T.-P. Weng, ``An adaptive sensor fusion method with applications
  in integrated navigation,'' \emph{The Journal of Navigation}, vol.~61, no.~4,
  pp. 705--721, Oct. 2008.

\bibitem{huang2017structure}
Y.-P. Huang, L.~Sithole, and T.-T. Lee, ``Structure from motion technique for
  scene detection using autonomous drone navigation,'' \emph{IEEE Transactions
  On Systems, Man, And Cybernetics: Systems}, Sep. 2017.

\bibitem{zanella2016best}
A.~Zanella, ``Best practice in {RSS} measurements and ranging,'' \emph{IEEE
  Communications Surveys \& Tutorials}, vol.~18, no.~4, pp. 2662--2686, Apr.
  2016.

\bibitem{niculescu2004positioning}
D.~Niculescu, ``Positioning in ad hoc sensor networks,'' \emph{IEEE Network},
  vol.~18, no.~4, pp. 24--29, Jul. 2004.

\bibitem{rahaim2012state}
M.~Rahaim, G.~B. Prince, and T.~D. Little, ``State estimation and motion
  tracking for spatially diverse {VLC} networks,'' in \emph{Global
  Communication Conference (GLOBECOM) Workshops}.\hskip 1em plus 0.5em minus
  0.4em\relax IEEE, Dec. 2012, pp. 1249--1253.

\bibitem{vatansever2017visible}
Z.~Vatansever and M.~Brandt-Pearce, ``Visible light positioning with diffusing
  lamps using an extended {Kalman} filter,'' in \emph{Wireless Communications
  and Networking Conference (WCNC)}.\hskip 1em plus 0.5em minus 0.4em\relax
  IEEE, Mar. 2017, pp. 1--6.

\bibitem{nguyen2014improvement}
N.-T. Nguyen, N.-H. Nguyen, V.-H. Nguyen, K.~Sripimanwat, and A.~Suebsomran,
  ``Improvement of the {VLC} localization method using the {Extended Kalman
  Filter},'' in \emph{IEEE Region 10 Conference (TENCON)}.\hskip 1em plus 0.5em
  minus 0.4em\relax IEEE, Oct. 2014, pp. 1--6.

\bibitem{nlom2017evaluation}
S.~M. Nlom, K.~Ouahada, A.~R. Ndjiongue, and H.~C. Ferreira, ``Evaluation of
  the {SFSK-OOK} integrated {PLC-VLC} system under the influence of sunlight,''
  in \emph{International Symposium on Networks, Computers and Communications
  (ISNCC)}.\hskip 1em plus 0.5em minus 0.4em\relax IEEE, May 2017, pp. 1--5.

\bibitem{allik2013ballistic}
B.~Allik, M.~Ilg, and R.~Zurakowski, ``Ballistic roll estimation using {EKF}
  frequency tracking and adaptive noise cancellation,'' \emph{IEEE Transactions
  on Aerospace and Electronic Systems}, vol.~49, no.~4, pp. 2546--2553, Oct.
  2013.

\bibitem{de2011uav}
H.~G. De~Marina, F.~J. Pereda, J.~M. Giron-Sierra, and F.~Espinosa, ``{UAV}
  attitude estimation using unscented {Kalman} filter and {TRIAD},'' \emph{IEEE
  Transactions on Industrial Electronics}, vol.~59, no.~11, pp. 4465--4474,
  Aug. 2011.

\bibitem{euston2008complementary}
M.~Euston, P.~Coote, R.~Mahony, J.~Kim, and T.~Hamel, ``A complementary filter
  for attitude estimation of a fixed-wing {UAV},'' in \emph{2008 IEEE/RSJ
  International Conference on Intelligent Robots and Systems}.\hskip 1em plus
  0.5em minus 0.4em\relax IEEE, Sep. 2008, pp. 340--345.

\bibitem{garcia2018design}
J.~Garcia~Guzman, L.~Prieto~Gonzalez, J.~Pajares~Redondo, S.~Sanz~Sanchez, and
  B.~Boada, ``Design of low-cost vehicle roll angle estimator based on {Kalman}
  filters and an {IoT} architecture,'' \emph{Sensors}, vol.~18, no.~6, p. 1800,
  Jun. 2018.

\end{thebibliography}



%


\end{document}